\documentstyle[preprint,prc,aps]{revtex}
\begin{document}
\draft
\title{Slow proton production in deep-inelastic neutrino scattering on
deuterium}
\author{G.D.~Bosveld, A.E.L.~Dieperink}
\address{Kernfysisch Versneller Instituut, NL-9747~AA~~Groningen,
         The Netherlands}
\author{A.G.~Tenner}
\address{NIKHEF-H, P.O.Box 41882, NL-1009 DB Amsterdam, The Netherlands}
 %
 %
\maketitle
\begin{abstract}
The cross section for semi-inclusive deep inelastic
charged current neutrino scattering
on hydrogen and deuterium in which a slow proton is observed in
coincidence with the muon is computed as a function of Bjorken $x$
and light-cone momentum of the detected proton.
In the Impulse Approximation
contributions from hadronization and (in case of the deuteron)
the emission of a spectator nucleon are considered. In addition
the probability for rescattering is computed.
The results are compared to a recent analysis of the data from the
WA25 (BEBC) experiment at CERN.
\end{abstract}
\section{Introduction}
Inclusive deep inelastic lepton scattering on nucleons is a well established
tool for the investigation of the quark-parton model. In the Bjorken limit the
cross section is proportional to a  structure function which depends on only
one scaling variable $x$, that can be interpreted as the light-cone momentum
fraction of the hit quark. Deep inelastic scattering on nuclear targets showed
\cite{SLOprp88} that the $x$ distribution of quarks in nucleons bound in a
nucleus differs from the one for a free nucleon (EMC effect). While part of
this effect can be explained  in terms of nuclear binding and fermi motion
\cite{BICjpg89,DIEprc91} there is evidence that additional possibly
non-nucleonic effects are required to obtain agreement with the data. However,
since in inclusive experiments one averages over all, strongly and weakly bound
nucleons, it is not possible to draw definite conclusions about the nature of
these effects.

More detailed information on the deep inelastic scattering process can be
obtained from more exclusive experiments. Here we concentrate on semi-inclusive
deep inelastic scattering (SIDIS) on the deuteron in which at least one slow
(i.e. low energy) proton is detected in coincidence with the scattered lepton.
There are several reasons why this process is of interest both on a free
nucleon and on nuclear targets.

On a free nucleon slow protons originate from hadronization of the spectator
quarks in the struck nucleon (so-called target fragmentation). In a previous
study \cite{BOSplb91,BOSprc92}  SIDIS on a free proton was considered, assuming
factorization of the cross section into a structure function $F(x)$ and a
spectator quark (debris) fragmentation function $D(z)$ , where $z$ is the
light-cone momentum fraction carried by the observed proton. It was found that
the ratio $P(x,z)$ of the SIDIS cross section $\sigma(x,z)$ and the inclusive
cross section $\sigma(x)$ has a strong bias for small $x$ values (in agreement
with experiment). Qualitatively this behavior can be explained by kinematics: a
proton at rest in the target rest system can only be produced at $x=0$; on the
other hand the dependence on $z$ and some details of $x$ dependence are
sensitive to the shape of the fragmentation function.

On nuclear targets there are several mechanisms that can give rise to slow
protons. First there is the above mentioned debris fragmentation of the quarks
in the struck bound nucleon, giving rise to creation of new particles and
possible change of momenta of spectator nucleons. From a comparison with the
free nucleon one obtains information on the effects of the nuclear medium on
the hadronization process.

Another category of slow protons in deep inelastic scattering on nuclei
are  spectator nucleons which are emitted from nucleon-nucleon correlations
present in target ground state. In the special case of pure two-body
correlations the l.c. momentum fraction $z$ of an observed spectator proton
directly determines the l.c. momentum of the nucleon being struck by the
lepton, $y \approx 2-z$. In this case one deals with the tagged EMC effect
(originally proposed by Frankfurt and Strikman \cite{FRAprp81}. This is of
interest since it might help to determine whether the convolution approach to
the EMC effect is appropriate \cite{DIEprc91,HELprc90}, or non-nucleonic
degrees of freedom are required \cite{FRAprl92,OELnpa90}.

Whereas in scattering on a free nucleon the direct hadronization process is the
only one possible, in nuclei both the direct and spectator processes occur and
it is of interest to investigate their characteristic behavior. In this paper
we concentrate on the simplest possible nucleus, the deuteron. We compute the
double differential cross section $\sigma(x,z)$ for the production of slow
protons with l.c. momentum fraction $z$. For the experimental data we use the
result of a recent new analysis of the WA25 collaboration. In the latter a
(partial) decomposition of the (anti-)neutrino SIDIS cross section is achieved
in terms of spectator, direct and rescattering events. In particular the
spectator contribution can be uniquely identified by sampling the slow protons
in the backward hemisphere. Thus it allows one to compare the various processes
separately with experimental results.

Another interesting feature  of the deuteron target is that it provides
information on the fragmentation of the neutron into a slow proton in
(anti-)neutrino induced reactions. In a pure valence quark approach the cross
section for $\nu+ n \rightarrow \mu+ p+ X$ would vanish. In practice one  must
also take into account sea-quark contributions (important  at small $x$). The
latter can be treated in two ways, namely as a 4-quark fragmentation
or as a bag of valence quarks surrounded by a pion cloud, which is struck by
the lepton. While these models give similar results for inclusive processes
there are interesting differences for exclusive reactions. For example,  in the
pion model one can produce slow protons in $ \nu+n $ in two ways: via
$n\rightarrow p+\pi^-$, and via $n\rightarrow\Delta^0+\pi^0$  with subsequent
decay of the isobar into $p+ \pi^-$, whereas in the naive leading order 4-quark
fragmentation only the latter process contributes to the production of slow
protons

Finally, in the present study we discuss possible effects of final state
interactions. After hadronization of the quarks in the hit nucleon (determined
by the formation time) the hadrons which are produced in forward direction can
in general interact with the spectator nucleons and thus lead to a possible
modification of their momentum distribution and charge and the creation of new
particles. The possibility to study the dependence of rescattering on the
formation (hadronization) time is of interest since it plays an essential role
in the interpretation of heavy ion reactions.

The paper is organized as follows. In section 2 we summarize the SIDIS cross
section on a free proton. The hadronization is described in terms of
fragmentation functions for di-quarks and four-quarks. In section 3  we
discuss the direct and spectator contribution for the deuteron assuming plane
wave impulse approximation and compare with experiment. The effect of
rescattering on the spectator momentum distribution is estimated in the eikonal
approximation and the role of the formation time is discussed. Section 4
contains concluding remarks.

\section{Slow proton production on a nucleon}
\subsection{Introduction}

In the Bjorken limit inclusive deep inelastic scattering (DIS) on a nucleon is
characterized (in leading order) by one scaling variable $x=Q^2/2p\cdot q$,
where $Q^2=-q^2$ is the square of the four-momentum transfer and $p$ the
nucleon momentum. In the parton model a DIS event at a specific $x$ corresponds
to the knock-out of a quark carrying a fraction $x=-q^-/p^-$ of the nucleon
light-cone momentum (where $q^\pm=q^0\pm q^3$ and  the momentum transfer
$\vec{q}$ is in the direction of $z$). In case of neutrinos the cross section
can be expressed in terms of three structure functions $F_i(x), \; (i=1,2,3)$
which only depend upon the quark momentum distributions.

The physical picture immediately after the (leptonic) scattering process is
that of one quark moving with a high momentum with respect to the debris of the
target (the quark cluster), cf. Fig.~\ref{Fig:ClusterFragmentation}.
Ultimately, the quark and the cluster will interact via the strong color field,
in which quark-antiquark pairs are produced, yielding  color neutral objects in
the final state.

In semi-inclusive deep inelastic scattering (SIDIS) one observes one or more
particles in coincidence with the scattered lepton.  For our purposes it is
convenient to distinguish between two limiting processes, namely the production
of mainly fast particles originating from the fragmentation of the struck
leading quark, and the production of mainly slow hadrons coming from
fragmentation of the debris.  As to the leading quark fragmentation one can
show \cite{LEVprep93} that in leading order (in $1/Q$) the cross section can be
factorized into a structure function $F(x)$ which describes the DIS on a quark
and a probability for hadronization.  The latter can be parametrized by a
fragmentation function $D_{\{q\}}^{\hbox{\rm\scriptsize x}}(z_f,p_\perp)$ with
the property that
 $D_{\{q\}}^{\hbox{\rm\scriptsize x}}(z_f,p_\perp)dz_fdp_\perp$
is the probability that a `jet' with
quark-content $\{q\}$ yields a fragment $\rm x$ carrying a fraction
$z_f=p^+_{\rm x}/q^+ $ of the initial quark light-cone momentum and transverse
momentum $p_{\perp}$.

The  fragmentation process of the quarks left behind (debris or target
fragmentation) is more complicated. In this case one cannot use perturbative
QCD arguments to prove factorization even in leading order \cite {LEVprep93}.
Still in practice one usually also assumes that the cross section factorizes.
The appropriate variable for target fragmentation is
 $z_s=p_{\rm x}^-/(p^-+q^-)$
which is related to the observable fraction of the target
momentum: $z=p_{\rm x}^-/p^-=(1-x)z_s$.

\subsection{Semi inclusive cross section}
The angle integrated semi-inclusive cross section for deep inelastic
(anti-)neutrino scattering on a nucleon with the observation of a hadron x with
l.c. momentum fraction $z_s$ and transverse momentum and $p_{\perp}$ can be
written as
\begin{equation}
   \frac{d^3\sigma^{\bar{\nu},\nu}}{dxdz_sdp_\perp}
 =
   \frac{G^2mE}{\pi}
   \left[ \frac{x}{3}{\cal F}_1+\frac12{\cal F}_2\pm\frac{x}{3}{\cal F}_3
   \right] \,,
\end{equation}
with $G$ the Fermi coupling constant and $E$ the energy of the (anti-)neutrino.
{}From now on we will only consider target fragmentation.
In the factorized approximation  the semi-inclusive proton structure functions
for neutrino scattering ${\cal F}_i^{\nu p}$ can be expressed as
\begin{eqnarray}
    {\cal F}_2^{\nu p}(x,z_s,p_\perp)
  & = &
    2x\left[ d(x)D_{\{p/{d}\}}^{\hbox{\rm\scriptsize x}}(z_s,p_\perp)
   +\bar{u}(x)D_{\{p/{\bar{u}}\}}^{\hbox{\rm\scriptsize x}}(z_s,p_\perp)\right]
    \label{Eq:FSIDIS2} \, ,\\
    {\cal F}_3^{\nu p}(x,z_s,p_\perp)
  & = &
    2\left[ d(x)D_{\{p/{d}\}}^{\hbox{\rm\scriptsize x}}(z_s,p_\perp)
   -\bar{u}(x)D_{\{p/{\bar{u}}\}}^{\hbox{\rm\scriptsize x}}(z_s,p_\perp)\right]
   \label{Eq:FSIDIS3} \, ,
\end{eqnarray}
and ${\cal F}_1={\cal F}_2/2x$. Here $d(x)$ and $\bar{u} (x)$ are the down-and
anti-up quark momentum distributions in the nucleon, respectively; $p/d$ and
$p/\bar{u}$ denote the cluster resulting after the knockout of a down and
anti-up quark from a proton, respectively. The structure functions for
anti-neutrinos are obtained from those of the neutrinos by replacing
$d\leftrightarrow\bar{d}$ and $u\leftrightarrow\bar{u}$.

Since the neutrino can interact with both valence and sea quarks, the number of
quarks in the debris (cluster) varies. When a valence quark is hit, a diquark
cluster remains, while the knockout of a sea quark (which is expected to be
important at small $x$ values) results in (at least) a four quark cluster. The
fragmentation functions for these processes are expected to have a different
dependence on $z$\cite{FRAprp88}. Hence it is convenient to decompose the
semi-inclusive cross sections for neutrino proton scattering into a valence and
a sea contribution,
$\sigma^{\nu p}=\sigma^{\nu p}_{\rm val}+\sigma^{\nu p}_{\rm sea},$ with
\begin{equation}
   \frac{d^3\sigma_{\rm val}^{\nu}}{dxdzdp_\perp}
  =
  \frac{G^2mE}{\pi}\frac{2x}{1-x}d_v(x)
  D_{\{uu\}}^p\bigl(\frac{z}{1-x},p_\perp\bigr),
\end{equation}
and
\begin{equation}
   \frac{d^3\sigma_{\rm sea}^{\nu}}{dxdzdp_\perp}
  =
  \frac{G^2mE}{\pi}\frac{2x}{1-x}\left[
   d_s(x)D_{\{p/{d}_s\}}^p\bigl(\frac{z}{1-x},p_\perp\bigr)
 +\frac13
   \bar{u}_s(x)D_{\{p/\bar{u}_s\}}^p\bigl(\frac{z}{1-x},p_\perp\bigr)
   \right] \, . \label{Eq:FactorizedSea}
\end{equation}
Here the subscript $\{p/q_s\}$ denotes the four-quark system $\{uud\bar{q}_s\}$
and we have neglected strange quarks. For slow fragments (i.e. $z\approx 1$)
the diquark fragmentation function $ D^{\hbox{\rm\scriptsize x}}_{\{qq\}} $ can
be approximated by a
splitting function \cite{BARprd82},  multiplied by combinatorial
flavor/isospin factors  $\gamma_{\{qq\}}^{\hbox{\rm\scriptsize x}}$ and  a
transverse momentum distribution
\begin{eqnarray}
 D_{\{qq\}}^{\hbox{\rm\scriptsize x}}(z_s,p_\perp)
 &=& \lambda_{qq} \gamma_{\{qq\}}^{\hbox{\rm\scriptsize x}} (z_s-z_{\rm min})
      \theta(z_s-z_{\rm min}) f(1-z_s)\rho(p_\perp) \, ,
\end{eqnarray}
where $z_{\rm min}$ (corresponding to a proton at rest in the hadronic
center-of-mass frame) is given by  $z_{\rm min}= \sqrt{m^2+p^2_{\perp}}/W$ with
$W$ the invariant mass of the final system, $W^2=m^2+Q^2(1/x-1)$
\cite{MELzpa92}. For the transverse momentum distribution we assume the form
\begin{eqnarray}
     \rho(p_\perp) &=& \frac{4p_\perp}{\left\langle p_\perp \right\rangle^2}
                        e^{-2p_\perp/\left\langle p_\perp \right\rangle} \, ,
\end{eqnarray}
with $\left\langle p_\perp^2 \right\rangle\approx 300\, (\hbox{\rm MeV}/c)^2$.
For the splitting function
$f$ we use the counting rule estimate $f(\xi)=\xi^n \,$ (with
$n=1)$\cite{BLAprd74,BROprd78}.
The factors $\gamma_{\{qq\}}^{\hbox{\rm\scriptsize x}}$ are given in
Table~\ref{Tab:Gammas}. The overall normalization $ \lambda_{qq}$ can be
obtained by requiring that the integral of $D$ over the appropriate $z$ range
is normalized to unity,
\begin{equation}
   \sum_{\rm x} \int_{z_{\rm min}}^1 dz_s D^{\rm x}_{{qq}} (z_s)=1 \,.
   \label{Eq:NormD2q}
\end{equation}

The fragmentation of the four-quark system (resulting from the interaction with
a sea-quark) has not been dealt with in any detail in the past. Ishii et al.
\cite{ISHplb89} assume a unit probability for fragmentation into a nucleon. In
the present study we need in addition to the $z$ dependence a prescription for
the relative branchings into $p,n$ and the various isobars. Since for the
later application to $A=2$ it is convenient to use a  simple form; in the
present paper we use the parametrization\cite{MELzpa92}
\begin{equation}\label{Eq:FourQuarkFragmentation}
   D_{\{4q\}}^{\hbox{\rm\scriptsize x}}(z_s)
  =
 \lambda_{4q}\gamma^{\hbox{\rm\scriptsize x}}_{\{4q\}}\theta(z_s)\theta(1-z_s)
 \, .
\end{equation}
We will consider two possible choices for $\gamma^{\hbox{\rm\scriptsize
x}}_{\{4q\}}$:
\par\noindent
(i) The most naive approach \cite{ISHplb89} is to assume that the produced
baryon has the same valence quark structure as the hit nucleon (i.e. $p\to
p,\Delta^+$ and $n\to n,\Delta^0$). The resulting
$\gamma^{\hbox{\rm\scriptsize x}}_{\{4q\}}$ are
listed in Table~\ref{Tab:Alphas}.
\par\noindent
(ii) Alternatively we can simulate the sea quarks by a meson cloud model. Such
a model has recently been used to describe the sea-contribution in SIDIS; it is
a very effective mechanism for the production of slow
nucleons\cite{MELzpa92,BOSprc92,KORzpa92}. In this case the coincidence cross
section  is given by\cite{KORzpa92}
\begin{equation}
   \frac{d^3\sigma^\nu_{\rm sea}}{dxdzdt}
  =
  \frac{G^2s}{2\pi}(1-z)\frac{g^2_{\pi NN}}{(4\pi)^2}\frac{t}{(t+m_\pi^2)^2}
  F^{\nu\pi}_2\left(\frac{x}{1-z}\right)|F_{\pi NN}(t)|^2 \, ,
\end{equation}
where $t=-(p_{\rm x}-p)^2$, $g_{\pi NN}$ is the pion nucleon coupling constant
and $F_{\pi NN}(t)$ the corresponding form factor.
(The corresponding expression for the
production of a Delta is given in Ref.~\cite{KORzpa92}.)

In practice it is
convenient to use the factorized form of Eqs.~(\ref{Eq:FactorizedSea}) and
(\ref{Eq:FourQuarkFragmentation}), retaining only the branching ratios
$\gamma_{\{4q\}}^{\rm x}$ that result from the meson cloud model.
These involve the different isospin couplings at the
$\pi NN$  and  $\pi N\Delta$ vertices (which determine  the relative magnitudes
of $g_{\pi NN}$ and $g_{\pi N\Delta}$),  the different couplings of the probe
with the  neutral and charged mesons and a parameter $\kappa$ which determines
the  relative ratio of nucleon and  isobar production. The latter is taken
equal to  one \cite{KORzpa92}. The resulting values for values for
$\gamma^{\hbox{\rm\scriptsize x}}_{\{4q\}}$  are listed in
Table~\ref{Tab:Sullivan}.  Analogously to
Eq.~(\ref{Eq:NormD2q}) the $\lambda_{4q}$ follows from  the  normalization of
the four-quark fragmentation function is such that
\begin{equation}
 \sum_{\rm x}\int_{z_{\rm min}}^1 dz_sD_{\{4q\}}^{\hbox{\rm\scriptsize x}}(z_s)
  = 1 \, .
\end{equation}
One noticeable difference between the above two prescriptions
is that in a $\nu +n$ interaction a proton can be produced in approach (ii)
but not in (i).

The probability for finding a slow proton in a momentum bin $\Delta p$ is
obtained by integrating the triple differential cross section over the
corresponding $(p_\perp,z)$ domain and dividing by the inclusive cross section
\begin{equation}
   P(x,\Delta p)
  =
   \int_{\Delta p}dzdp_{\perp} \sigma(x,z,p_{\perp}) / \sigma(x) \, .
\end{equation}
Note that, apart from the direct production of protons, there is also a
significant contribution from decay of nucleon isobars. This is taken into
account by assuming that the resonances decay isotropically in their rest frame
\cite{KORzpa92,BOSprc92}.

In Fig.~\ref{Fig:Proton} we show the ratio $P(x,\Delta p)$ of the calculated
SIDIS cross section and the inclusive cross section as a function of $x$ for
(anti-)neutrino  scattering on the free proton  compared with the data of Guy
et al. \cite{GUYplb89}. Since in that paper only a relative  (`renormalized')
cross section  was given we normalized the published data to our calculated
cross section in the
low bin for $\nu+ p$.

For comparison in Fig.~\ref{Fig:NeutronVsExp}  the neutron
the experimental cross section as extracted from the deuterium experiment (as
discussed in the next section) is shown.
\par\noindent
The following points are worth noting:
\par\noindent
(i) Whereas the proton data have an arbitrary normalization, it  should be
emphasized that the neutron data are absolute.
\par\noindent
(ii) Qualitatively the fall off of $P(x, \Delta p)$ with $x$
can be ascribed to kinematics: slow protons correspond to $z \approx 1$ while
$z+x<1$. Only the details of the dependence on $x$ are sensitive to
 the precise form of
the fragmentation function. The presence of an experimental tail both in
the $\nu+p$ and $\nu+n$ data which extends beyond the kinematical
limit, $x=1-z$, is puzzling and cannot be explained. The larger tail in the
proton data may be due to the absence of an unsmearing of $x$ in the analysis.
(In ref. \cite{MELzpa92} this taken into account by
an (ad hoc) smearing of the $x$ distribution.)
\par\noindent
(iii) The ratio of $P(x,\Delta p)$ for the
low and high bin, which depends on the form of the fragmentation
function near $z=1$, is somewhat underestimated; the sea
contribution dominates at small $x$ values and is more
important  for the low bin than in the high bin.
\par\noindent
(iv) Experimentally, $P(x,\Delta p)$ for $\nu$ and $\bar{\nu}$ scattering on a
neutron are quite different, and almost equal for a proton target. For the
$\bar{\nu} +n$ process the valence contribution almost vanishes, and the cross
section is strongly affected by sea-quark effects;
the sea contribution is larger when using the pionic model than naive
fragmentation.
\par\noindent
(v) Comparing the different choices for the 4q fragmentation function one sees
that in $\bar{\nu}+p$ the naive 4q fragmentation  yields a larger
cross section than the pion model  while in $\nu+ n$  one has the opposite
situation.
This  is a consequence of the presence of an important
 charged meson exchange contribution in the pion cloud approach.
\par\noindent
(vi) The unfavoured fragmentation process $\bar{\nu}+n \rightarrow \mu^++p+X$
which does not have a direct contribution
proceeds only via an intermediate $\Delta^0$ after scattering
on the sea-quarks in the pion cloud approach. The experimental cross section
does not show such a dramatic effect.
\section{The Deuteron}

\subsection{Classification of events} \label{Sec:Spectator}

In a nuclear medium there are several processes that can give rise to slow
protons. First in analogy with the DIS on a free nucleon target slow protons
can originate from  fragmentation of the spectator quarks (debris
fragmentation) in the struck nucleon with the other nucleons acting  as
spectators. This  process will be referred to as the {\it direct} process.  A
new effect, which only occurs for $A>1$, is the deep inelastic scattering off a
bound nucleon  and the observation of a spectator proton (called the {\it
spectator} process). Since in a mean field description the $A-1$ nucleus is
bound, this process provides evidence for the presence of correlations in the
ground state (for $A>2$). The important point is that in the independent pair
approximation neglecting the centre-of-mass motion of the pair and final state
interactions one has a direct relation between the l.c. momentum fractions of
the detected ($z$) and struck ($y$) nucleons $y+z=2$.

In the following we will compute the SIDIS cross section on the deuteron with a
slow proton present assuming that the above processes contribute incoherently.
An additional process that will be considered separately is the possible
reinteraction ({\it rescattering})  of the hadrons produced in the
fragmentation with the spectator nucleon.  The latter process is of interest
since it depends on the hadronization in the nuclear medium, which may lead to
a modification of the momentum and possibly the charge of the observed
spectator nucleon.

The data to be used as a comparison were obtained from an analysis of deep
inelastic (anti-)neutrino scattering experiment on the deuteron in a bubble
chamber in which all (charged) particles were detected \cite{BEBC}.
The experimental data allow for a separation into `even' and `odd' events,
corresponding to an even or odd number of charged hadrons in the final state,
respectively. From charge conservation it follows that odd events correspond
to the direct process on the neutron, where the spectator proton has
insufficient momentum to become visible in the chamber. On the other hand,
the category of even events has different constituents. First, there is the
direct process on the proton, leaving the neutron as an invisible spectator.
Second, there is the direct process on the neutron where the proton
spectator is visible. And third, there are primary reactions on either a proton
or a neutron with a subsequent rescattering against the other nucleon.
Spectator events may be individually identified when the slow proton is
emitted into the backward hemisphere. The direct process on a target at rest
only yields secondary protons in the forward hemisphere. (For the small
contamination of direct protons, emitted backward by fermi motion, a
correction is made.)
{}From the events with a backward spectator, events
with a forward spectator may be constructed.

In addition, a selection may be made of rescattering events by means of
kinematical arguments. The quantity
\begin{equation}
    \epsilon = \sum_f (E_f - p_f^3) - m
\end{equation}
is evaluated, where the sum runs over all visible particles of the event.
If all particles are visible and correctly identified, $\epsilon = 0$,
otherwise $\epsilon < 0$. On the other hand, for rescattering events
where the target is the deuteron rather than a single nucleon with mass $m$,
$\epsilon$ may become larger than 0. A more detailed description of the
experimental analysis of rescattering is found in
Refs.~\cite{TENrep86,TENrep88,TENinc92}.

\subsection{Direct process}

To compute the cross section for the direct process we use a convolution model
\cite{BOSplb91,BOSprc92}, in which the fermi motion and binding are taken into
account by folding the semi-inclusive structure functions for a free nucleon
with the momentum distribution of the nucleon in the nucleus. In this approach
off-shell effects and possible final state interactions between the quarks in
the struck nucleon and the spectator nucleons are neglected.

When using the fragmentation approach for the scattering off valence and sea
quarks the cross section can be expressed as an integral over the light-cone
momentum fraction carried by the nucleon (cf.~Fig.~\ref{Fig:SemDisProcess}a):
\begin{eqnarray}
    \frac{d^4 \sigma^{\nu d}_{\rm dir}(x,z)}{dxdzd^2\vec{p}_\perp}
 &=&
    \frac{G^2mE_\nu}{\pi}
    \frac{1}{1-x}
    \sum_{\tau =p,n} \int_{x+z}^{m_d/m}
    dyd^2\vec{k}_\perp f_d(y,\vec{k}_\perp)\hfill\nonumber \\
  & & \hskip-2.5cm  \left\{
     \frac23{\cal F}_2^{\nu\tau}
     \left(\frac{x}{y}
          ,\frac{z}{y-x},|\vec{p}_\perp-\frac{z}{y}\vec{k}_\perp|\right)
    +\frac{x}{3y}{\cal F}_3^{\nu\tau}\left(\frac{x}{y}
          ,\frac{z}{y-x},|\vec{p}_\perp-\frac{z}{y}\vec{k}_\perp|\right)
    \right\}\, , \label{Eq:DirectCrossSection}
\end{eqnarray}
where the structure functions are defined in Eqs.~(\ref{Eq:FSIDIS2}) and
(\ref{Eq:FSIDIS3}) and  the generalized convolution function is given by
\cite{OELnpa90,BOSprc92}
\begin{equation}
 f(y,\vec{p}_{\perp})
                    = my\int\frac{d^3k}{\sqrt{k^2+m^2}} n(k)
                     \delta\bigl(y- \frac{m_d-\sqrt{k^2+m^2}-k_3}{m}\bigr)
                     \delta^{(2)}(\vec{k}_\perp-\vec{p}_\perp)\, .
    \label{Eq:ConvFuncPt}
\end{equation}
After integration over the transverse momentum one recovers the usual
convolution function for the light-cone momentum \cite{BOSprc92}
\begin{equation}
     f_d(y)
   =
     2\pi m^2y
     \int_{k_{\rm min}}^{\infty}\!\!\frac{kdk}{\sqrt{k^2+m^2}}\,n(k) \,,
    \label{Eq:ConvfuncDeuteron}
\end{equation}
where $k_{\rm min} = \left|\frac{(my-m_d)^2-m^2}{2(m_d-my)}\right|$.
The factor $y$ in Eq.~(\ref{Eq:ConvFuncPt})
which comes from the convolution of the hadronic tensor, is usually referred to
as the flux factor.
Alternatively one can convolute the semi-inclusive cross section;
in this case the factor $y$ arises from the presence of
the invariant mass of the neutrino plus nucleon system in the
convolution integral (Ref.~\cite{FRAprp81}).

In the present work the momentum density $n(k)$ is taken from
Ref.~\cite{MACprp87} (Bonn potential). For the deuteron the  convolution
function $f_d(y)$ is sharply peaked at $y=1$, corresponding to a  nucleon at
rest, and falls of rapidly for $y\neq 1$ and as a result the cross section for
the direct process differs very little from the one for the free nucleon case.
The main difference is that, due to the fermi-motion, the calculated cross
section extends to larger values of $x$.   Furthermore, the number of protons
produced in the backward direction with a momentum above the experimental
cutoff of $150\, \hbox{\rm MeV}/c$ is negligible.

As noted above it is possible to compare the results for the direct
cross section on the neutron with the data set for odd events, to which the
even events with a backward and a constructed forward spectator are added.
The calculated $P(x,\Delta p)$ as a function of $x$ for the free neutron
(shown in Fig.~\ref{Fig:NeutronVsExp})
and those for the bound neutron (shown in
Fig.~\ref{Fig:OddVSmodel})
differ slightly both in the value of the cut-off in $x$
(due to the fermi motion in the deuteron ) and in normalization (due
to the different inclusive cross sections for $A=1$ and $A=2$).
Again the smaller SIDIS cross section for the $\bar{\nu}$ case can be
explained by the presence of the unfavoured fragmentation function.

\subsection{Spectator process}

In the absence of final state interactions the cross  section for the
spectator process is given by (cf.~Fig.~\ref{Fig:SemDisProcess}b)
\begin{eqnarray}
  \frac{d^4\sigma_{spec}^{\nu d}}{dxdzd^2\vec{p}_{\perp}}
  & = &
   \frac{G^2mE_\nu}{\pi}
   \int_x^{m_d/m}dy
   \left\{
     \frac23F_2^{\nu n}\left(\frac{x}{y}\right)
    +\frac{x}{3y}F_3^{\nu n}\left(\frac{x}{y}\right)
   \right\}
  \nonumber \\
   &  &
   f_d(\frac{m_d}{m}-z,\vec{p}_{\perp})\delta(z+y-m_d/m) \,.
\label{Eq:SpectatorCrossSection}
\end{eqnarray}

Hence the dependence of cross section on $z$ (when integrated over $p_{\perp}$)
is determined by the light-cone momentum fraction of the neutron in the
deuteron (approximately symmetric around $z=1$) and the corresponding shift in
the arguments of the structure functions (which is asymmetric around $z=1$).
One sees that the maximum l.c. momentum fraction of the spectator for given $x$
is given by $z_{\rm max}=m_d/m-x$ corresponding to a  backward momentum
\cite{CARplb91} $p_{\rm max} = \frac{m^2-(m_d-xm)^2}{2(m_d-xm)}$.  In
particular, for $x=0$ one has $p_{\rm max}=\frac34m$.
Because the convolution function is strongly peaked at $y=1$ the spectator
cross
section decreases rapidly when going away from $z=1$.

In Fig.~\ref{Fig:SpectatorVSmodel} the calculated ratio of the semi-inclusive
and inclusive cross sections $P(x,\Delta p)$ for the spectator process is
compared with the backward spectator spectrum as extracted from the even hadron
set.  $P(x,\Delta p)$ depends only weakly on $x$, since the scattering process
on the neutron is essentially inclusive. The  difference in $x$ dependence
between neutrinos and anti-neutrinos can be attributed to the difference of the
$x$ distributions between the $u$ and $d$ quarks in the proton and the neutron:
spectator protons correspond to events where the $\nu \; (\bar{\nu})$ scatters
off a $d \; (u)$ quark in the neutron whereas the inclusive deuteron cross
section involves an average over proton and neutron events.

Also of interest is the  ratio $R(z,\Delta x)$ obtained by integrating the
semi-inclusive cross section over the transverse momentum and $x$, and
dividing by the inclusive cross section
\begin{equation}
   R(z,\Delta x)
  =
   \int_{\Delta x}dxdp_{\perp} \sigma(x,z,p_{\perp})
  /\int_{\Delta x}dx\sigma(x) \, .
\end{equation}
Fig.~\ref{Fig:XBinnedSpectator} shows
that $R(z,\Delta x)$ for the spectator process (only backwards), sharply peaked
at $z=1$, is well reproduced. (The plateau around $z=1$ is due to the lower
momentum cutoff at $150$ MeV/c.)

Finally, in Fig.~\ref{Fig:ThreeDimPlot} we show the calculated ratio for the
sum of direct and spectator cross sections.
The peak at $z=1$ corresponds to detection of the spectator,
whereas the ridge at small values of $x$ stems from the direct production of
protons.

\subsection{Rescattering}

Above we have restricted ourselves to the PWIA, i.e. neglected effects from
final state interactions between the spectator and the  hadrons produced by the
fragmentation. In the present context two aspects of fsi are of interest. First
as shown in \cite{TENinc92} the inclusive rescattering probability (which
experimentally can be identified  with the help of the variable $\epsilon$) is
sensitive to the time (length) scale in the hadronization process. A second
topic of interest is the modification of the spectator l.c. momentum
distribution.

The deuteron is the simplest nuclear system where the hadronization process in
a medium  can be studied, and the nuclear structure can be treated without
approximation. Below we discuss the consequences of final state interactions
(rescattering) for both the spectator spectrum and the rescattering
probability.

\subsubsection{Inclusive rescattering probability}

The simplest approach to rescattering is to assume that the hit nucleon
hadronizes instantaneously into a number of color singlets which can rescatter
incoherently on the spectator. It was shown in \cite{TENinc92} that  in this
independent rescattering model the observed rescattering fraction $P_{\rm
resc}$ (the number of rescatter events over the total number of events) is
grossly overestimated. In the eikonal approximation the incoherent rescattering
probability is\cite{BENprl92}:
\begin{equation}
  P_{\rm resc}(x,Q^2)=\frac{N(x,Q^2)\sigma}{4\pi}
                  {\left\langle\frac{1}{r^2}\right\rangle_d} \, ,
  \label{Eq:RescatterFractionIncoherent}
\end{equation}
where $\sigma$ is the fragment-spectator cross section and $N(x,Q^2)$ is the
multiplicity.  The latter can be parametrized as
\cite{BELprd79,CHAprl76}
\begin{equation}
   N(x,Q^2)=1.1+1.35\ln{W^2(x,Q^2)}
   \label{Eq:InvariantMass}\,.
\end{equation}
where $W(x,Q^2)$ is the invariant mass of the hadronic system:
\begin{equation}
   W(x,Q^2)=\sqrt{m_N^2+Q^2(1/x-1)}\, .
\end{equation}
This would predict too large a magnitude of $P_{\rm resc}$ and moreover, a
logarithmic increase in the rescattering probability as a function of the
energy transfer $\nu\sim Q^2/x$.  In contrast, the experiment\cite{TENinc92}
shows a saturation for large $\nu$ and a very small $x$ dependence
(cf.~Fig.~\ref{Fig:RescatterProbability}).

An alternative to the incoherent rescattering model is to replace
$N(x,Q^2)$ by $N_{\rm eff} =1$, independent of $x$.  Using for the
fragment-nucleon cross section $\sigma$ a value of $45 \rm mb$ one obtains
$P_{\rm resc} \approx 0.12 $. This agrees with the average experimental value.
This suggests that effectively only one of the produced colorless
fragments will interact.  Alternatively, this can be viewed as the rescattering
of an object (the exited nucleon or string) with the transverse size of the hit
nucleon moving through the nucleus.

As shown in \cite{TENinc92} the most plausible explanation of
the small rescattering probability is the finite formation time
for the newly created particles. We have estimated this effect
by means of a Monte Carlo calculation and use the VENUS hadronization
code (which is based on string fragmentation) to generate a space-time spectrum
of fragments. To simulate coherence we assume that only the fragment which is
formed closest to the scattering centre will rescatter. Introducing
the distribution $F_{\hbox{\rm\scriptsize x}}(x,r_0)$ which gives the
probability that the
first fragment  is of type ${\hbox{\rm x}}$ and is formed at a distance $r_0$
from the scattering center,
the rescattering probability can be expressed as \cite{BENprl92}:
\begin{equation}
  P_{\rm resc}(x) =
  \int_0^\infty dr_0\, F_{\hbox{\rm\scriptsize x}}(x,r_0)\int_{r_0}^\infty dr\,
    \frac{\sigma\rho_d(r)}{4\pi r^2} \, .
  \label{Eq:RescatterFractionFormation}
\end{equation}
In Fig.~\ref{Fig:RescatterProbability} the inclusive rescattering probability
as given by Eq.~(\ref{Eq:RescatterFractionFormation}) is plotted as a function
of $x$ (where we have folded in the experimental $Q^2$ distribution).  Also
shown is the rescattering probability for the independent rescattering model
with $N=1$.  One sees that this simple model for rescattering does not explain
the data: the rescattering probability is underestimated and the $x$ dependence
is not present in the data.

This strongly indicates that the present approach is
too simple: we have neglected all interactions of the string (excited nucleon)
before hadronization and assumed that only one fragment will rescatter.  Any
interaction of the string will however, increase the rescattering probability
and smoothen the $x$-dependence.

\subsubsection{Spectator modification}

The spectator process has been proposed as a means to measure correlations in
nuclei. In the independent pair approximation where two nucleons move back to
back in the nucleus in their cm system after striking one nucleon its partner
can leave the nucleus with its original lc momentum if fsi are neglected. In
the backward direction one thus measures the nucleon lc momentum distribution.
There are several ways to study whether this simple picture is affected by fsi.
A simple criterium is the ratio  of $x$ values for events with and without a
slow backward proton, $v_z=\langle x\rangle_{\rm bp}/\langle x\rangle$, which
in the case of  spectator events is proportional to $2-z$. In case of $^{20}$Ne
this relation has been confirmed qualitatively for events  containing only one
slow proton\cite{MATzpc89}.

The most obvious contribution to the rescattering process, that gives rise to
slow protons, is that of the elastic
rescattering of fragments on the spectator. The resulting events are thus
removed from the spectator and added to the rescattering sample. The spectrum
of rescattered spectators can be calculated using a generalization of
$F_{\hbox{\rm\scriptsize x}}(x,r_0)$ which includes the distribution of the
longitudinal momentum
$p_3$ carried by the fragments\cite{BOSinprep}.

Using this distribution we can write the modified backward proton spectrum as
as a sum of rescattered and undistorted spectators
\begin{equation}
   \sigma^{\rm bp}=\sigma^{\rm resc}+\tilde\sigma^{\rm spec} \, ,
\end{equation}
where $\tilde\sigma^{\rm spec}=(1-P_{\rm resc})\sigma^{\rm spec}$ is the cross
section for spectators, that are unaffected by rescattering, and
$\sigma^{\rm resc}$ the spectrum of protons that arises from the rescattering
process.

As a first estimate of the effect of rescattering we calculated the distorted
spectator distribution (including only elastic rescattering). For the cross
section we used the parametrization from Ref.~\cite{ZAVzpc86} for all types of
fragments (including resonances and  mesons).
In Fig.~\ref{Fig:RescatteredSpectators} we plot the spectrum of backward
protons
\begin{equation}
   R(z,\Delta x)=\left(\int dx \frac{d\sigma^{\rm bp}}{dxdz}\right)
   \left(\int dx \frac{d\sigma^{\rm deut}}{dx}\right)^{-1}
\end{equation}
as a function of $z$. It is seen that, as one expects, the spectrum is shifted
towards the forward hemisphere ($z<1$). However, because of the rather small
effect, and the number of approximations in the calculation  it is not possible
to make a quantitative comparison with the (anti-)neutrino data.

\section{Concluding remarks}

Motivated by a reanalysis of experimental data we have investigated the
absolute cross section for semi-inclusive deep inelastic charged current (anti)
neutrino  scattering on hydrogen and deuterium with the observation of a slow
proton in the final state, as a function of the Bjorken $x$ and the light-cone
momentum fraction $z$.

On a free nucleon, we assume that the cross section can be factorized into a
quark momentum distribution and a fragmentation function for the spectator
quarks. At small values of $x$ the cross section is dominated by sea-quark
effects; the latter can be simulated by a pion cloud model, and by a four-quark
fragmentation function. The strong bias of the ratio $P(x,z)$ of the
coincidence cross section and the inclusive cross section for small $x$ values
has basically a kinematic origin.

The deuteron target offers several new interesting effects. First, the data
analysis makes it possible to isolate deep inelastic scattering on the neutron
with subsequent fragmentation into a proton. Second, there is an additional
source of slow protons in the final state, namely the emitted spectator
nucleon. In the backward hemisphere where this process dominates, the cross
section is directly proportional to the light-cone momentum distribution of the
nucleons. In most cases qualitative agreement with the data is obtained; a more
quantitative comparison can only be made using higher accuracy data. Finally,
an analysis of rescattering events indicates that the naive incoherent
rescattering of produced hadrons is not consistent with experiment.

There are several reasons why it is interesting to extend the present study to
heavier nuclei. In the plane wave impulse approximation  the spectator momentum
distribution depends sensitively on details of the nuclear spectral function
$P(k,E)$. In addition one expects to obtain more information
about final state interactions between spectator nucleons and produced hadrons,
in which the formation time plays a crucial role.

\section*{Acknowledgments}

The authors thank the Amsterdam-Bergen-Bologna-Padova-Pisa-Saclay-Torino
Collaboration for supplying the experimental data.  They thank the staff of the
CERN SPS and the bubble chamber BEBC for establishing the experiment.  The
authors wish to thank O.~Scholten, C.~Korpa, M.~Strikman, L.~Frankfurt,
P.J.~Mulders and J.~Guy for stimulating discussions.  This work was supported
by the Foundation for Fundamental Research (FOM) and the Netherlands
Organization for the Advancement of Scientific Research (NWO).

\begin{figure}
\caption[ ]{Debris fragmentation in DIS} \label{Fig:ClusterFragmentation}
\end{figure}
\begin{figure}
\caption[ ]{
        Ratio $P(x,\Delta p)$ of the semi-inclusive production cross section
        of slow protons ($\nu(\bar\nu)+p\rightarrow\mu+p+X$)
        off free protons and the corresponding
        total inclusive cross section for two momentum bins of the
        slow proton (left 150-350 MeV/c, right 350-600 MeV/c). For the
        calculation the branching ratios from Tables \protect\ref{Tab:Alphas}
        (set 1) and
        \protect\ref{Tab:Sullivan} (set 2) have been used.
        Comparison is made with the data from
        Ref.~\protect\cite{GUYplb89} which have been normalized to
        the the calculated ratio for the low bin in neutrino scattering.}
\label{Fig:Proton}
\end{figure}
\begin{figure}
\caption[ ]{Ratio $P(x,\Delta p)$ for slow proton production off the neutron,
        equivalent to Fig.~\ref{Fig:Proton}.
        A comparison is made with experimental
        deuterium data as explained in Sec.~3. No normalization has been
        performed, the theoretical curves act as absolute predicions of the
        data.}
\label{Fig:NeutronVsExp}
\end{figure}
\begin{figure}
\caption{Direct (a) and spectator process (b)
for semi-inclusive deep-inelastic scattering on $A=2$.}
\label{Fig:SemDisProcess}
\end{figure}
\begin{figure}
\caption{The calculated ratio
$P(x,\Delta p)$ for the direct process on the neutron
 bound in the deuteron compared to the data.}
\label{Fig:OddVSmodel}
\end{figure}
\begin{figure}
\caption{The ratio
         $P(x,\Delta p)$ for the spectator process in the deuteron compared to
         the data (backward spectators only).}
\label{Fig:SpectatorVSmodel}
\end{figure}
\begin{figure}
\caption{The ratio
         $R(z,\Delta x)$
    for the spectator process as a function of $z$ in two
   different $x$ bins compared to the data.}
\label{Fig:XBinnedSpectator}
\end{figure}
\begin{figure}
\caption{Calculated ratio
        $R(x,z)=\sigma(x,z)/\sigma(x)$ of sum of the cross sections for the
        spectator and  direct process in $\nu+d\to\mu^-+p+X$ and the
        inclusive cross section $\nu+d\to\mu^-+X$, with a lower momentum cutoff
        of 150 MeV/c for the observed proton.}
\label{Fig:ThreeDimPlot}
\end{figure}
\begin{figure}
\caption{Rescattering probabilities as a function of $x$, calculated using the
independent rescattering model with $N_{\rm eff}=1$ (dashed lines) and the
formation time model (solid lines), compared to the data.}
\label{Fig:RescatterProbability}
\end{figure}
\begin{figure}
\caption{The ratio $R(z,\Delta x)\,\,(0\!<\!x\!<\!0.1)$ for backward protons
calculated without (solid line) and with rescattering (dashed line).}
\label{Fig:RescatteredSpectators}
\end{figure}
\begin{table}
\caption[ ]{Relative weights $\gamma_{\{qq\}}^{\hbox{\rm\scriptsize x}}$
  for diquark fragmentation functions $D^{\hbox{\rm\scriptsize x}}_{\{qq\}}$.}
\label{Tab:Gammas}
\begin{center}
\begin{tabular}{ccccccc}
 $\{qq\}$&$\Delta^{++}$&$\Delta^+$&$\Delta^0$&$\Delta^-$&$p$&$n$ \\ \hline
 $uu$&$\frac{18}{27}$  &$\frac{6}{27}$  &  $0$   & $0$    &$\frac{3}{27}$
& $0$     \\
 $ud$&  $0$       &$\frac{2}{18}$  &$\frac{2}{18}$ & $0$    &$\frac{7}{18}$
&$\frac{7}{18}$  \\
 $dd$&  $0$       & $0$     &$\frac{6}{27}$ &$\frac{18}{27}$& $0$
&$\frac{3}{27}$  \\ \hline
\end{tabular}
\end{center}
\end{table}
\begin{table}
\caption[ ]{Relative weights $\gamma^{\hbox{\rm\scriptsize x}}_{\{4q\}}$ for
naive four-quark fragmentation}
\label{Tab:Alphas}
\begin{center}
\begin{tabular}{cccccccc}
  process&$\{4q\}$  &$\Delta^{++}$&$\Delta^+$&$\Delta^0$&$\Delta^-$&$p$&$n$\\
\hline
  $\nu p$&$p/{d}_s,p/\bar{u}_s$
             & $0        $&$1/2       $&$0$&$0$&$1/2$&$0$\\
  $\bar{\nu}p$&$p/{u}_s,p/\bar{d}_s$
             & $0        $&$1/2       $&$0$&$0$&$1/2$&$0$\\
  $\nu n$&$n/{d}_s,n/\bar{u}_s$
             & $   0     $ &$0         $&$1/2$&$0$&$0$&$1/2$ \\
  $\bar{\nu}n$&$n/{u}_s,n/\bar{d}_s$
             & $   0     $ &$0         $&$1/2$&$0$&$0$&$1/2$ \\ \hline
\end{tabular}
\centerline{\small The notation $\{p/d_s\}$ denotes the debris after hitting a
$d_s$ in the proton.}
\end{center}
\end{table}
\begin{table}
\caption[ ]{Relative weights $\gamma^{\hbox{\rm\scriptsize x}}_{\{4q\}}$ using
isospin branchings
 of the pion cloud model.}
\label{Tab:Sullivan}
\begin{center}
\begin{tabular}{cccccccc}
process &$\{4q\}$&$\Delta^{++}  $&$\Delta^+$     & $ \Delta^0$
 &$\Delta^-$     &$p  $        &$  n  $         \\ \hline
$\nu p$ &$p/d_s,p/\bar{u}_s$
        &$\frac{3\kappa}{4\kappa+1}$ &$\frac{\kappa}{4\kappa+1}$
& $ 0       $     &$    0   $     &$\frac{1}{4\kappa+1}$&$ 0   $         \\
$\bar{\nu} p$ & $p/{u}_s,p/\bar{d}_s$
        &$   0         $&$\frac{\kappa}{2\kappa+5}$
& $\frac{\kappa}{2\kappa+5}$   &$    0   $
&$\frac{1}{2\kappa+5}$&$\frac{4}{2\kappa+5}$   \\
$\nu n$ &$n/d_s,n/\bar{u}_s$
        &$   0         $&$\frac{\kappa}{2\kappa+5}$
& $\frac{\kappa}{2\kappa+5}$   &$    0   $
&$\frac{4}{2\kappa+5}$&$\frac{1}{2\kappa+5}$   \\
$\bar{\nu} n$ &$n/u_s,n/\bar{d}_s$
        &$   0         $&$  0  $        & $\frac{\kappa}{4\kappa+1}$
&$\frac{3\kappa}{4\kappa+1}$ &$ 0 $       &$\frac{1}{4\kappa+1}$   \\ \hline
\end{tabular}
\end{center}
\end{table}
\end{document}